\documentclass[journal=aaembp,manuscript=article,layout=twocolumn]{achemso}

\usepackage{soul}
\usepackage{graphicx}
\usepackage{dcolumn}
\usepackage{bm}% bold math
\usepackage{xcolor}
\usepackage{amssymb}
\usepackage{subfigure}
\usepackage{mathptmx}
\usepackage{amsmath}
\usepackage{mciteplus}
\usepackage{float}

\author{Juan P. Mendez}
\email{jpmende@sandia.gov}
\affiliation{Sandia National Laboratories, 1515 Eubank SE, Albuquerque, NM 87123, USA}

\author{Denis Mamaluy}
\email{mamaluy@sandia.gov}
\affiliation{Sandia National Laboratories, 1515 Eubank SE, Albuquerque, NM 87123, USA}

\title{Quantum charge sensing using a semiconductor device based on $\delta$-layer tunnel junctions}

\begin{document}

\begin{abstract}
We report a nanoscale device concept based on a highly doped $\delta$-layer tunnel junction embedded in a semiconductor for charge sensing. Recent advances in Atomic Precision Advanced Manufacturing (APAM) processes have enabled the fabrication of devices based on quasi-2D, highly conductive, highly doped regions, known as $\delta$-layers, in semiconductor materials. In this work, we demonstrate that APAM $\delta$-layer tunnel junctions are ultrasensitive to the presence of charges near the tunnel junction, allowing the use of these devices for detecting charges by observing changes in the electrical current. We demonstrate that these devices can enhance the sensitivity in the limit, i.e., for small concentrations of charges, exhibiting significantly superior sensitivity compared to traditional FET-based sensors. We also propose that the extreme sensitivity arises from the strong quantization of the conduction band in these highly-confined systems.
\end{abstract}

\maketitle

\section{Introduction}\label{sec:introduction}

Sensors are indispensable components in our lives nowadays. Among all existing types of sensors, field effect transistor (FET)-based sensors offer significant advantages for sensing\cite{Dai:2022}: they enable label-free electrical detection, provide real-time detection, can be small in size and weight, and, most importantly, can be integrated on a chip. FET-based sensors have been widely reported for various applications, including the detection of cortisol, a hormone released in response to stress \cite{Sheibani2021}, the SARS-CoV-2 virus associated with the COVID-19 pandemic \cite{Seo:2020}, nucleic acids such as DNA or RNA \cite{Hwang:2020}, prostate-speciﬁc antigen (PSA), a key protein marker for prostate cancer \cite{Gao:2010}, real-time detection of toxins in flowing water \cite{Maity:2023}, as well as the measurement of pH\cite{Gao:2010} and strain \cite{Lee:2019}, among other applications.

\begin{figure*}
  \centering
  \includegraphics[width=0.8\linewidth]{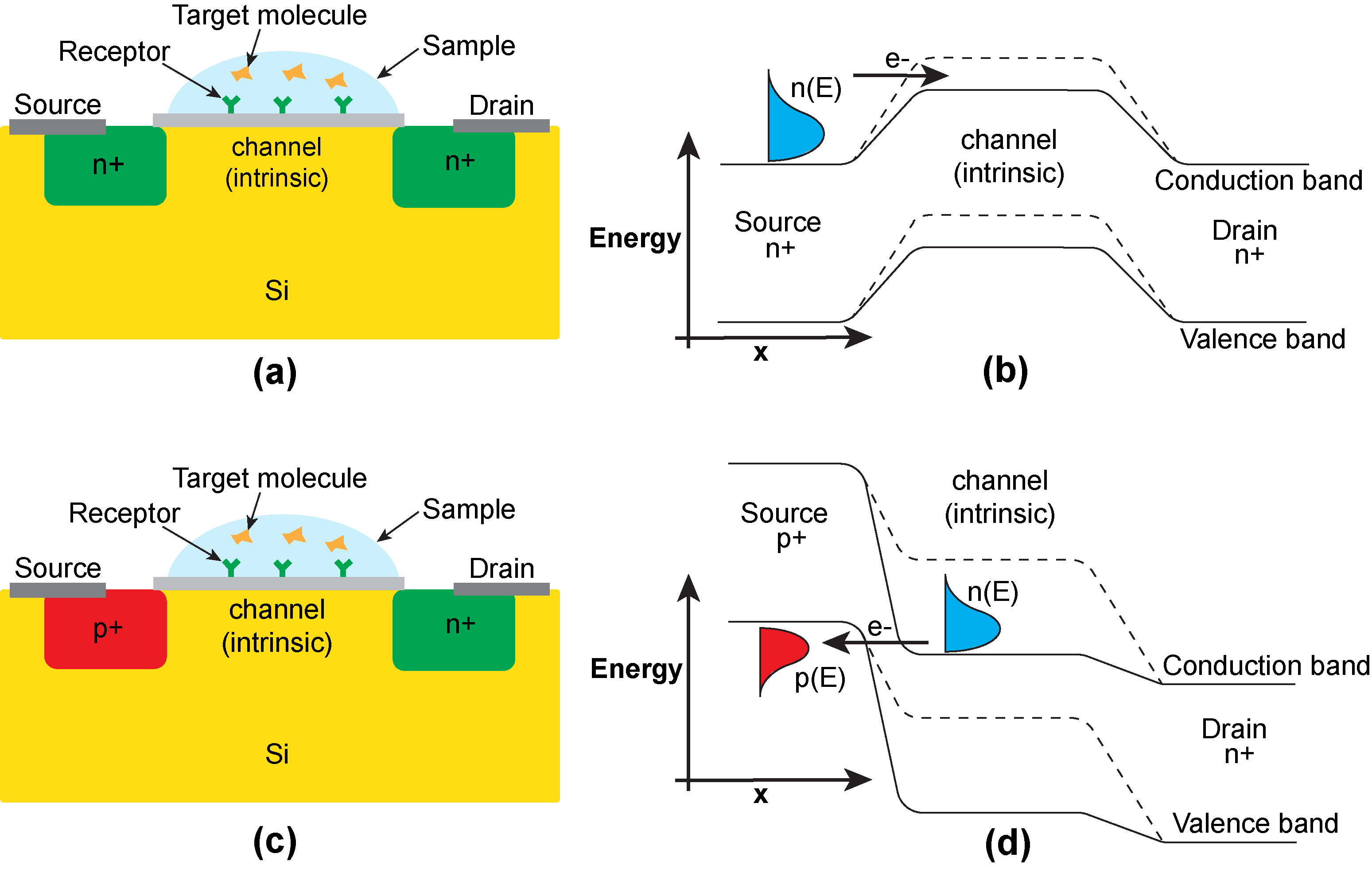}
  \caption{(a) Schematic representation of a convectional FET-based sensor. (b) A typical band structure for a convectional FET-based sensor. (c) Schematic representation of a tunnel FET-based sensor  ( TFET-based sensor). (d) A typical band structure for a TFET-based sensor.}
  \label{fig:CFET and TFET-based sensors}
\end{figure*}

The operation of conventional FET-based sensors is similar to that of a MOSFET. FET-based sensors consist of a source, drain, channel, insulator, and a sensing area, instead of the gate in MOSFETs. For example, in FET-based sensors for biological applications, the sensing area can be functionalized with receptor molecules that can trap the target molecules that we want to detect (Figure~\ref{fig:CFET and TFET-based sensors}a). When the target molecules attach to the receptors, they act as a gate or, alternatively, they apply an effective voltage, depleting the channel and allowing electrons to flow from source to drain (Figure~\ref{fig:CFET and TFET-based sensors}b). This results in an increase in current between the source and drain, with the magnitude of the current change being related to the concentration of the target molecules in the sample. However, FET-based sensors suffer from sensitivity limitations, i.e., they require a minimum concentration of the target molecules in the sample to produce a measurable change in current. The sensitivity can be enhanced using tunnel FET (TFET)-based sensors (Figure~\ref{fig:CFET and TFET-based sensors}c). In this case, when the target molecules attach to the sensing surface, it results in a lowering of the band structure of the channel (Figure~\ref{fig:CFET and TFET-based sensors}d). The signal, a measurable current change, is only received once the conduction band edge is lower than the valence band edge, enabling the band-to-band tunneling. Despite the enhanced sensitivity compared to their counterparts, the sensitivity still remains very low for low concentrations of the target molecules in the sample, as it still requires a minimum number of molecules to be bound to the sensing area to enable detectable band-to-band tunneling.

The sensitivity for (T)FET-based sensors operating in the subthreshold regime can be approximated as $S \approx 10^{\Delta \phi/ SS}-1$, where $\Delta \phi$ is the change of the surface potential on the sensing area induced by the presence of the element to be sensed, and $SS$ is the subthreshold swing of the device\cite{Sarkar:2012}. For a nanowire (T)FET based biosensor\cite{Gao:2010}, the change of the surface potential is related to the change of the charge on the surface by $\Delta Q = C \cdot L \cdot \Delta \phi$, where $C= (1/C_{ox}+1/C_{NW})^{-1}+C_{DL}$ is the capacitance per unit length between the surface charge and the nanowire/electrolyte systems, $C_{DL}$ is the electrolyte double layer capacitance per unit length, $C_{OX}$ is the oxide layer capacitance per unit length, $C_{NW}$ is the capacitance of the nanowire per unit length, and $L$ is the channel length. In the subthreshold regime, where the maximum sensitivity is reached, $C \approx C_{DL}$. It is not surprising, then, that the best value of $S$ for a nanowire TFET-based sensor to sense a single charge is just $S=8.0\times 10^{-3}$, assuming $SS=20$~mV/dec, $C_{DL}=4.7\times 10^{-15}$~F/um\cite{Gao:2010}, and $L=0.5$~um, which is also in agreement with previously theoretical sensitivity for TFET-based sensors reported in Ref.~\citenum{Sarkar:2012}. We propose that novel quantum-based sensors can significantly enhance the sensitivity for detection at the limit, i.e., reaching a sensitivity value of $S>>8.0\times 10^{-3}$ for very low charge concentration, and, consequently, opening new opportunities for biological, chemical, radiation, and nuclear sensing. 

\begin{figure*}
  \centering
  \includegraphics[width=1\linewidth]{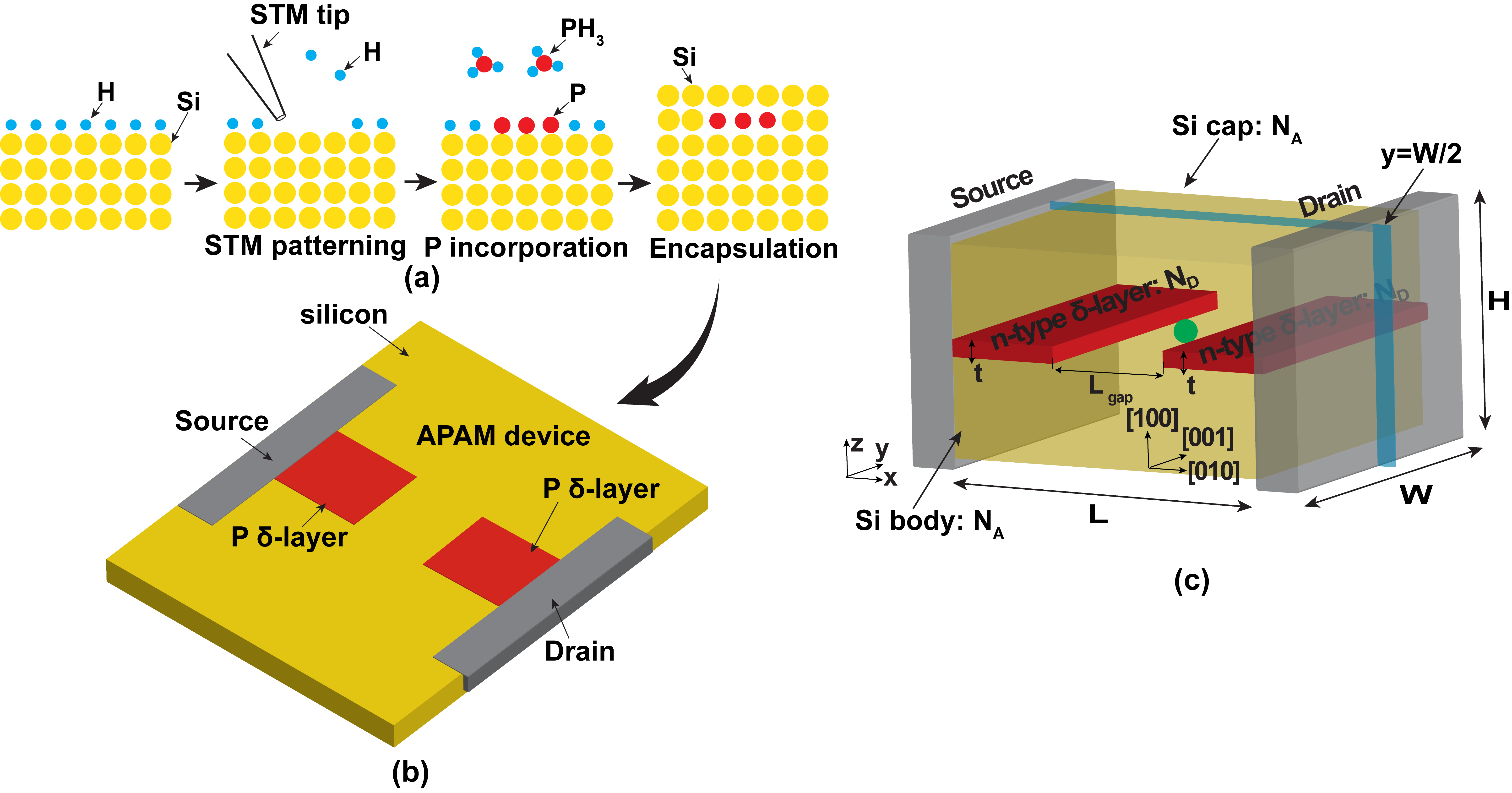}
  \caption{(a) Key steps in the Atomic Precision Manufacturing (APAM) process. (b) Example of a $\delta$-layer tunnel junction device, composed of two very thin, highly n-type-doped layers separated by an intrinsic semiconductor gap and embedded in silicon, fabricated using APAM. (c) Schematic of the computational device (APAM $\delta$-layer tunnel junction) used in our simulations.}
  \label{fig:APAM and computational_model}
\end{figure*}

Recent advances in Atomic Precision Advanced Manufacturing (APAM) have enabled the creation of quasi-2D highly-doped regions, also known as $\delta$-layers, in a semiconductor with single-atom precision \cite{Wilson:2006, Warschkow:2016, Fuechsle:2012, Ivie:2021b, Wyrick:2022} and high conductivity \cite{Goh:2006, Weber:2012, McKibbin:2013, Keizer:2015, vskerevn:2020, Dwyer:2021}. APAM is a manufacturing process used to incorporate dopants, such as P or B, at the atomic scale onto a Si surface using surface chemistry \cite{Ward:2020,vskerevn:2020}. This process involves several steps, as shown in Figure~\ref{fig:APAM and computational_model}a in a simplified way, to fabricate APAM devices such as the $\delta$-layer tunnel junction device shown in Figure~\ref{fig:APAM and computational_model}b. For P-doped $\delta$-layers embedded in silicon (Si: P $\delta$-layer ) \cite{Wyrick:2019,Ward:2020}, the process starts with the Si (100) surface fully passivated with H; using the tip of a Scanning Tunneling Microscope (STM), the $\delta$-layer structures are patterned by precisely removing H atoms from the exact locations where dopants will be incorporated; the Si surface is then exposed to a precursor gas, which contains the dopants, such as phosphene (PH$_3$) for P dopants\cite{Ward:2020}, followed by an annealing process to incorporate the dopants into the surface; finally, an epitaxial Si is overgrown through a series of annealing processes to protect the planar structure and activate the dopants.  APAM has various applications, including the exploration of novel electronic devices such as nanoscale diodes and transistors for classical computing and sensing systems \cite{Mahapatra:2011, House:2014, vskerevn:2020, Donnelly:2021}, as well as the exploration of dopant-based qubits in silicon for quantum computing\cite{Fricke:2021,Wang:2022}. In Ref.~\citenum{House:2014}, the sensitive detection of single charges using a planar tunnel junction has been evaluated experimentally at cryogenic temperatures, where it has been demonstrated that the junction conductance strongly responds to the electrostatic field of a gate and to electron transitions in a quantum dot. 

In this work, we propose a two-terminal APAM $\delta$-layer tunnel junction (Figure~\ref{fig:APAM and computational_model}b) as an ultra-sensitive device for charge sensing and elucidate the underlying mechanism responsible for the enhanced sensitivity. These devices can also offer significant advantages such as (i) they are exceptionally simple two-terminal devices; (ii) they are extremely small-size, on the order of 10~nm; and (iii) they are suitable for integration with CMOS technology. We demonstrate in this work that APAM $\delta$-layer tunnel junctions can easily detect single charges near the tunnel junction by observing measurable changes in the electrical current. We also show theoretically that these devices exhibit sensitivity superior to that of (T)FET-based sensors in the low-charge concentration limit. Finally, we propose that the extreme sensitivity arises from the strong quantization of the conduction band in these highly-confined $\delta$-layers.

\begin{figure*}[t]
  \centering
  \includegraphics[width=1\linewidth]{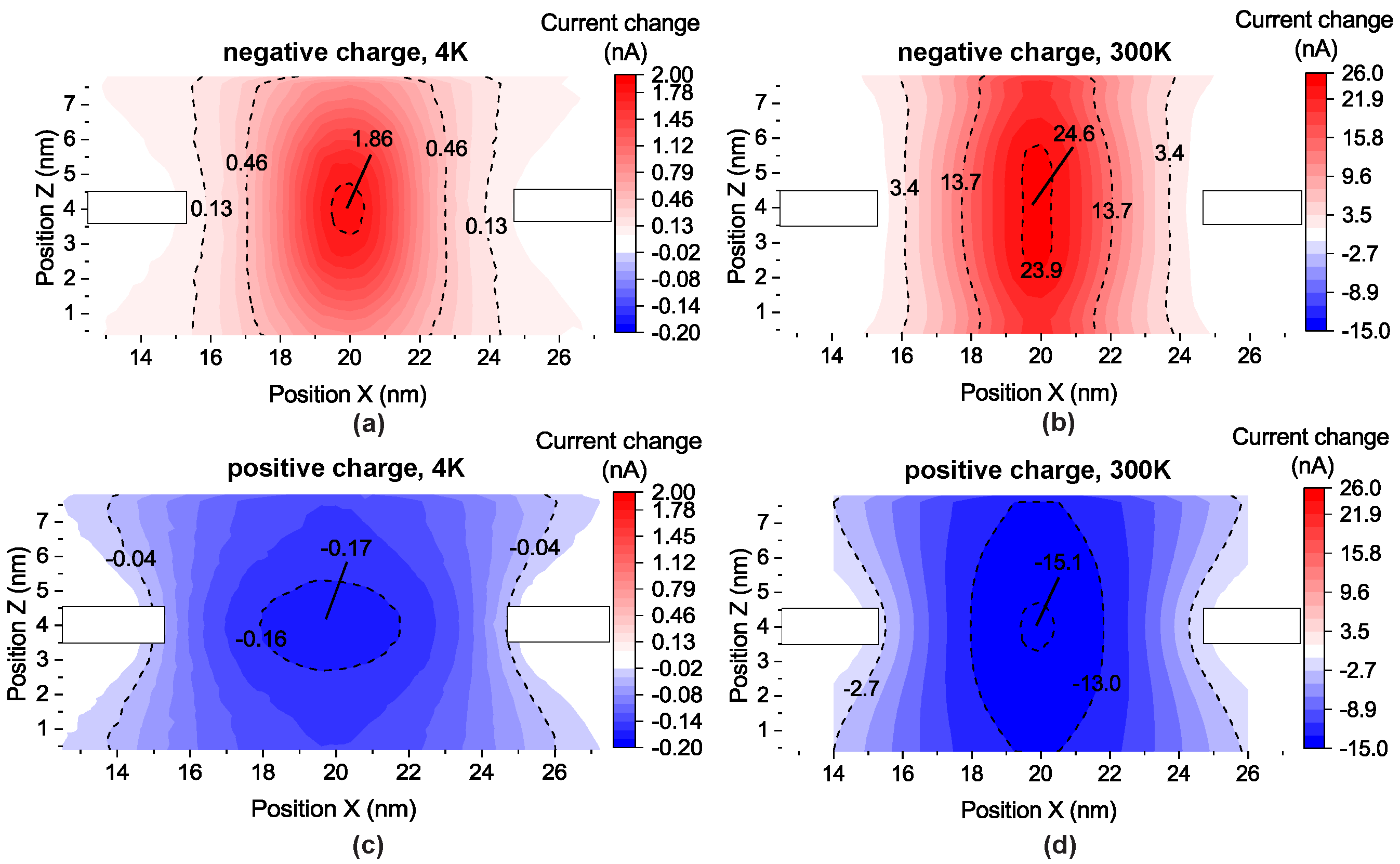}
  \caption{Contour maps of the tunneling current change due to the presence of a single negative charge at different positions (x,z) in the middle plane of the device (y=W/2) at 4~K in (a) and at 300~K in (b), as well as for a single positive charge at 4~K in (c) and at 300~K in (d). A voltage of 1~mV is applied between the source and drain. The nominal tunneling current (i.e., without the presence of charges near the tunnel junction) is around 0.27~nA at 4~K and 28.5~nA at 300~K.}
  \label{fig:current,1mV}
\end{figure*}

\section{Results and discussion}\label{sec:Results and discussion}

The first question we seek to address is whether $\delta$-layer tunnel junctions exhibit sensitivity to the presence of a minimal number of charges near the tunnel junction. In order to answer this question, we first check if the tunneling current changes when a single (negative/positive) charge is present near the tunnel junction. For our simulations, we adopt the structure of the $\delta$-layer tunnel junction shown in Figure~\ref{fig:APAM and computational_model}c, which consists of two very conductive, thin, highly n-type-doped (e.g., P) layers separated by an intrinsic semiconductor gap and embedded in silicon. We consider a $\delta$-layer with a thickness of $t=1.0$~nm and a donor sheet doping density of $N_D=1.0 \times 10^{14}$~cm$^{-2}$ ($N_D^{(2D)} = t \times N_D^{(3D)}$), and a tunnel junction length of $L_{gap}=10$~nm. The Si body and Si cap are lightly doped with acceptors, with a doping density of $N_A=5.0 \times 10^{17}$~cm$^{-3}$. Further details of the computational simulations can be found in the Methods section.

\begin{figure*}[t]
  \centering
  \includegraphics[width=1\textwidth]{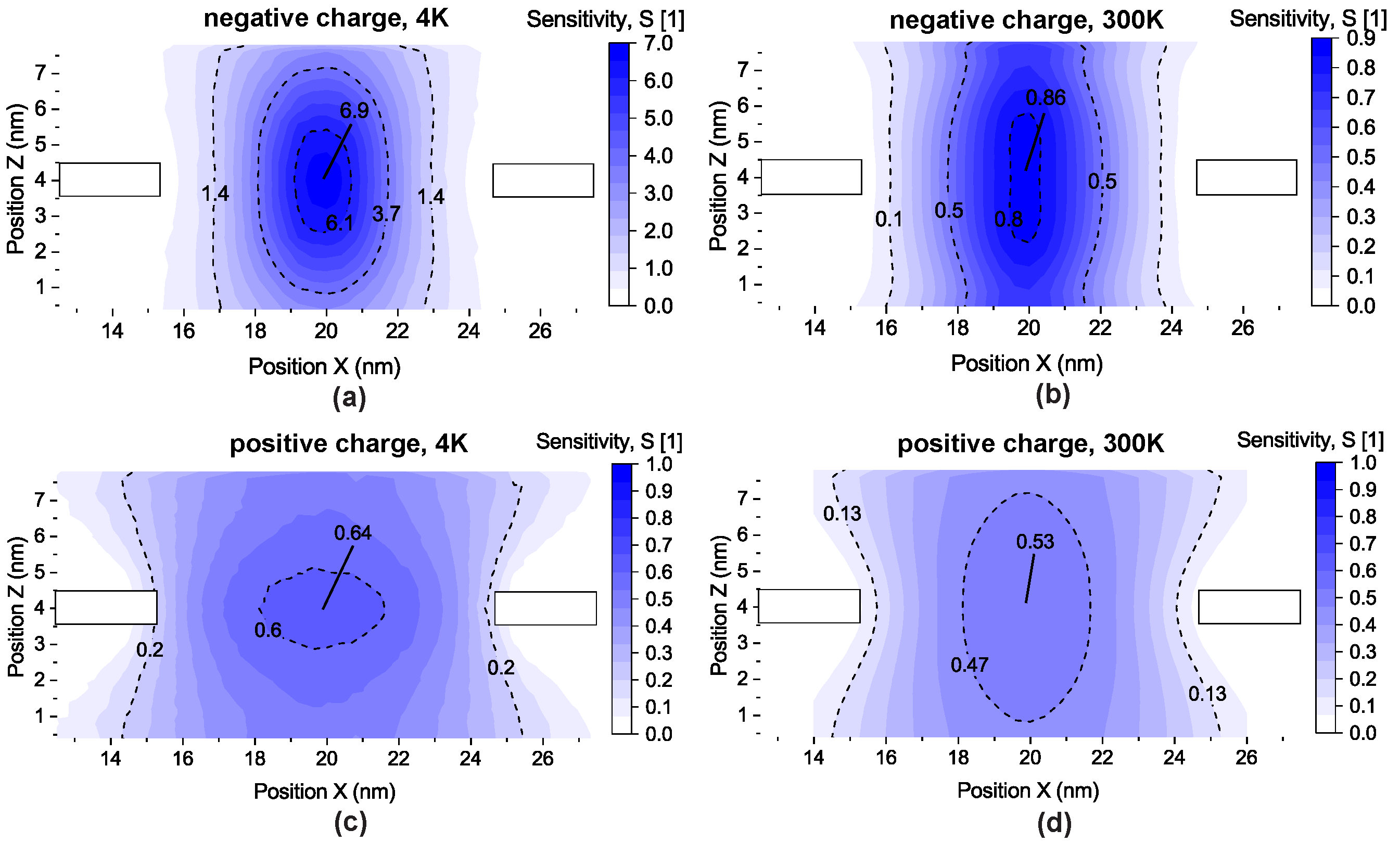}
  \caption{Sensitivity of the device due to the presence of a single negative charge located at different positions (x,z) in the middle plane of the device (y=W/2) at 4~K in (a) and at 300~K in (b), as well as for a single positive charge at 4~K in (c) and at 300~K in (d). A voltage of 1~mV is applied between the source and drain.}
  \label{fig:sensitivity,1mV}
\end{figure*}

Figure~\ref{fig:current,1mV} shows the tunneling-current contour map when a positive/negative charge is located at different positions (x,z) in the middle plane of the device (y=W/2), shown in Figure~\ref{fig:APAM and computational_model}c, at 4 and 300~K, when a voltage of 1~mV is applied between the source and drain.  These results indicate that the presence of either a negative/positive charge near the tunnel junction indeed strongly affects the tunnel rate in $\delta$-layer tunnel junctions. A negative charge increases the tunneling current, while a positive charge decreases it, demonstrating that our APAM device is indeed suitable for charge sensing. 
The maximum current change occurs when the charge is present in the center of the tunnel junction: at 4~K, with a decrease of up to 0.17~nA for a positive charge and an increase of up to 1.86~nA for a negative charge; at room temperature, with a decrease of up to 15.1~nA for a positive charge and an increase of up to 24.6~nA for a negative charge. However, most importantly, the tunneling rate not only changes when the charge is located between the $\delta$-layers, but also when it is located outside the $\delta$-layers, which is an important result if we want to use this device for charge sensing. 
These results also suggest that the impact of the electrical charge on the current diminishes as the charge moves farther away from the tunnel junction. Later in this work, we address how far the device can sense the presence of a charge by investigating how the change in current varies with the distance of the charge from the device.

\begin{figure*}[ht!]
  \centering
  \includegraphics[width=1.0\linewidth]{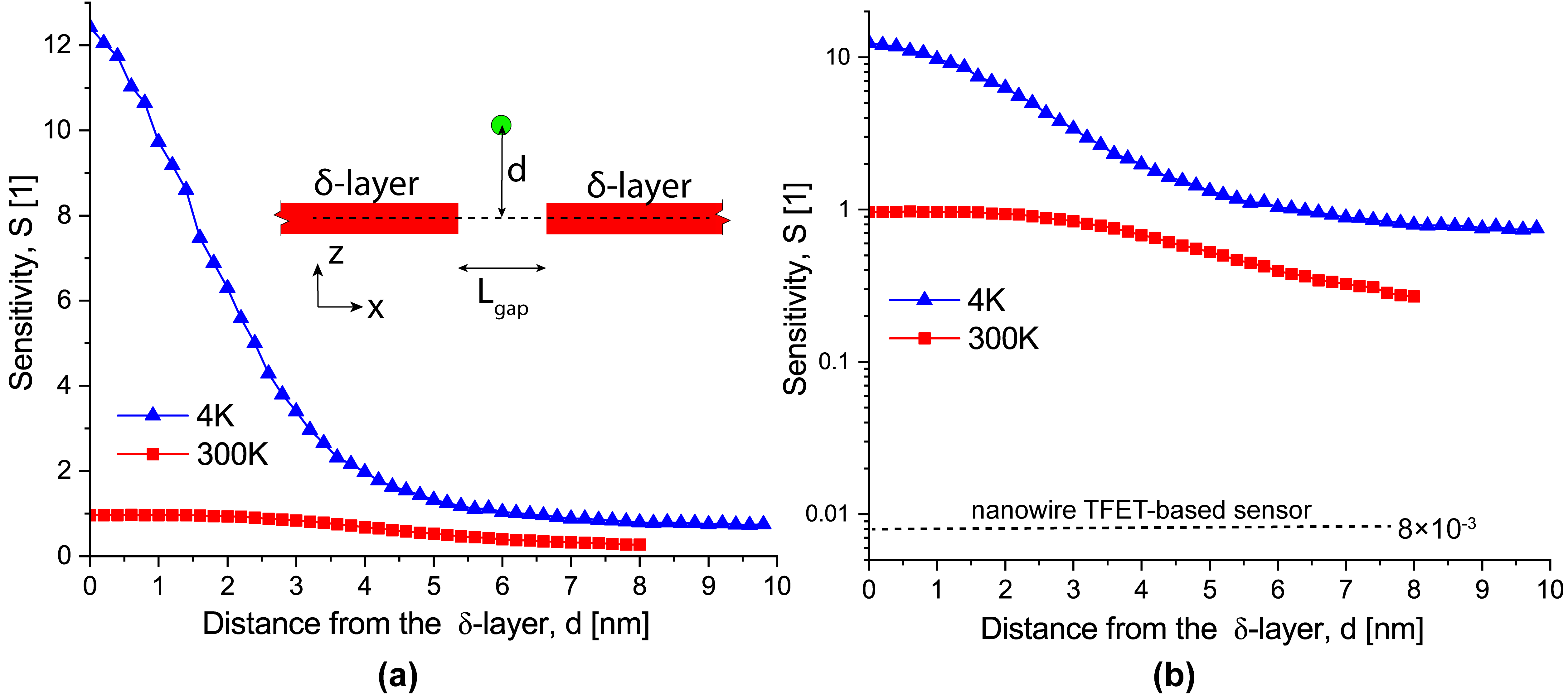}
  \caption{Sensitivity of the device to the presence of a single negative charge located at different distances $d$ from the $\delta$-layers, along the line $(x, y, z)=(L/2, W/2, d)$, at both temperatures 4 and 300~K: (a) Linear scale and (b) Log scale. The inset in (a) displays the schematic representation of a tunnel junction and a charge located at a distance $d$ from the $\delta$-layer. A voltage of 1~mV is applied between the source and drain. The device dimensions are: $L=40$~nm, $W=10$~nm, $H=15$~nm, and $t=1$~nm.}
  \label{fig:distance study}
\end{figure*}

Figure~\ref{fig:sensitivity,1mV} illustrates the sensitivity of the device. The sensitivity, $S$, defined as the ratio of the change in current caused by the presence of  a single charge to the nominal current, is given by $S=I_{\text{with charge}}/I_{\text{nominal}}-1$, where $I_{\text{with charge}}$ is the current when a single charge is present near the tunnel junction and $I_{\text{nominal}}$ is the current in the absence of charges near the tunnel junction. This metric indicates how much the tunneling rate changes relative to the nominal current of the device.  Figure~\ref{fig:sensitivity,1mV} indicates that the device exhibits very high sensitivity, especially for negative charges at low temperatures. At 4~K, we obtain a maximum sensitivity of 6.9 for sensing a negative charge in the middle of the tunnel junction, while the sensitivity for a positive charge is around 0.64. Similarly, the sensitivity decreases as the charge moves farther away from the center of the tunnel junction.
As expected, these results also show that the sensitivity decreases with the increase in temperature due to the contribution of the thermionic emission. At room temperature, we obtain a maximum sensitivity of $0.86$ for sensing a negative charge in the middle of the tunnel junction, while the sensitivity for a positive charge is $0.53$. There are two contributions to the total current in these systems: the thermionic emission 
($j_{\text{thermionic}}=AT^2\exp(-E_b/k_bT)$, where $E_b$ is the barrier energy) and the tunneling current.
The former corresponds to high-kinetic-energy electrons, with energies higher than the barrier energy at the junction, which can pass through tunnel junction without tunneling; the latter corresponds to electrons with energies lower than the barrier energy that can pass through the barrier via tunneling. If we approximate the sensitivity as $S=\Delta I_{\text{due to charges}}/(I_{\text{thermionic}}+I_{\text{tunneling}})$, where $I_{\text{thermionic}}$ and $I_{\text{tunneling}}$ are the nominal current contributions due to the thermionic emission and tunneling, respectively,  and $\Delta I_{\text{due to charges}}$ is the current change due to the presence of charges, we observe that sensitivity can rapidly decrease with temperature as $I_{\text{thermionic}}$ increases with temperature. 
Despite the decrease in sensitivity with temperature, these results also demonstrate that our APAM device achieves superior sensitivity compared to (T)FET-based sensors even at room temperature. The maximum sensitivity of our device is around $S=0.86$ at 300~K, which is significantly larger than the theoretical sensitivity of TFET sensors, typically $S \approx 8.0\times10^{-3}$ for low concentrations of charges. 

To achieve higher sensitivity for positive charges, comparable to that of negative charges, p-type $\delta$-layers can be used instead of n-type $\delta$-layers. Additionally, we also note that because the effects on the current due to the presence of a negative charge and a positive charge are dissimilar in magnitude, this device is also suitable for infrared photon detection, wherein the impact of a photon on the sensing area of the device results in a cascade of electron-hole pairs. An electron-hole pair cloud can be modeled as a single dipole with total charge $Q$ and a dipole moment $l$. In our previous work\cite{Mendez:2023}, we found that dipoles with sufficiently large moments can significantly affect the tunneling current in $\delta$-layer tunnel junctions, depending on the dipole orientation with respect to the $\delta$-layer system.

\begin{figure*}[ht!]
  \centering
  \includegraphics[width=1.0\linewidth]{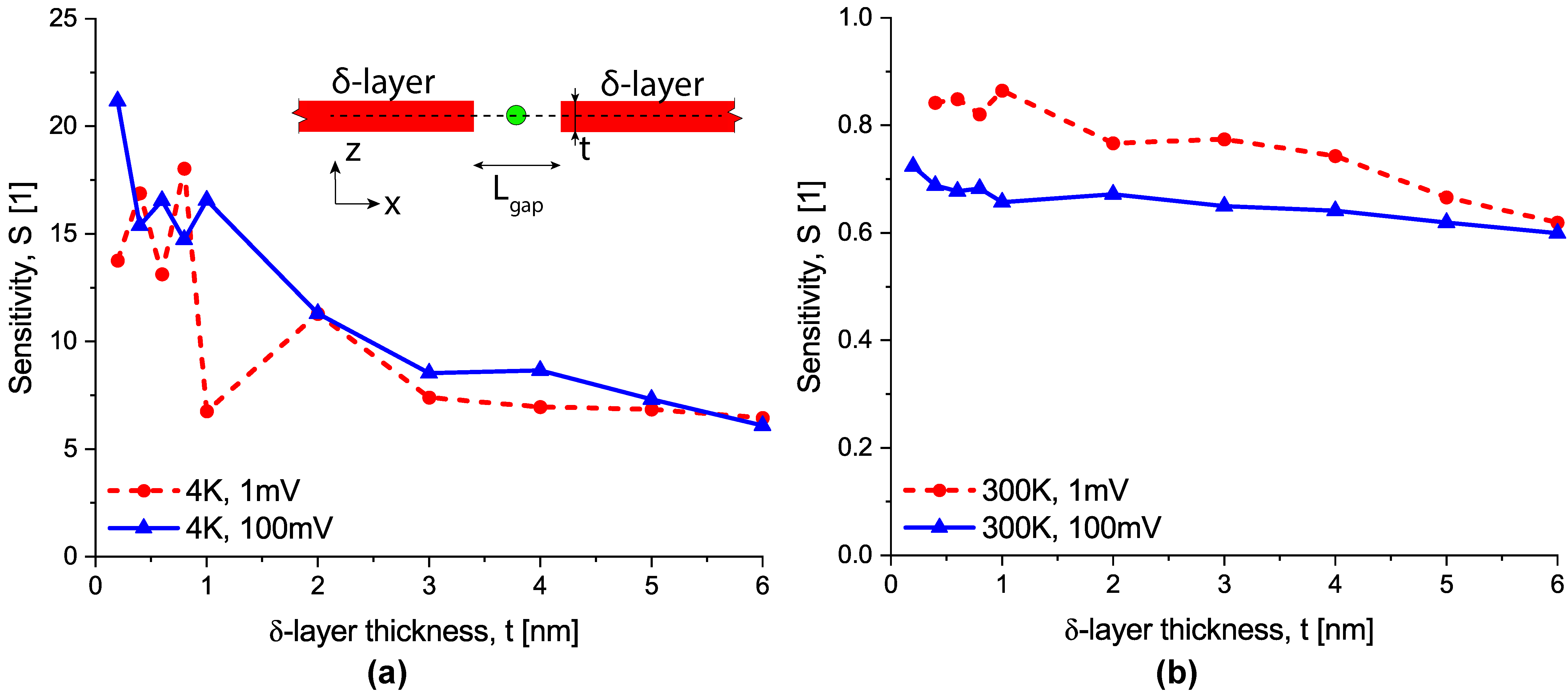}
  \caption{Study of sensitivity vs. $\delta$-layer thickness: (a) 4~K and (b) 300~K. The inset in (a) displays the schematic representation of a tunnel junction and a charge located at the center between the $\delta$-layers.}
  \label{fig:sensitivity vs thickness}
\end{figure*}

We next investigate how far the proposed device can sense the presence of charges. Figure~\ref{fig:distance study} shows the sensitivity of the device for different positions of a single negative charge located at a distance $d$ from the tunnel junction (see the inset in Figure~\ref{fig:distance study}a) at both temperatures, 4 and 300~K. The applied voltage between the source and drain is 1~mV. Due to limitations in the computational resources, in these simulations, we use a smaller device width of $W=10$~nm, while adopting a larger device height of $H=15$~nm, with the $\delta$-layers positioned asymmetrically 5~nm from the bottom of the device. At 4~K, the results indicate that the sensitivity initially decays nearly-exponentially with the distance from the $\delta$-layer tunnel junction at first, but then, at distances greater than 5~nm, the sensitivity starts reducing at a much slower rate, maintaining a relatively high value $S\sim1$. At 300~K, the sensitivity is roughly constant up to a distance of 3~nm, then it decays nearly-exponentially with the distance. Interestingly, the rate of the sensitivity decay  at short distances is higher at 4~K than at 300~K, while it is the opposite at long distances. Thus, from these results, we can conclude that $\delta$-layer tunnel junctions can detect charges over longer distances at lower temperatures, specifically at 4~K. We also note that narrower devices with stronger confinement exhibit greater sensitivity, specifically at 4~K, as we can conclude by comparing the results from Figures~\ref{fig:sensitivity,1mV} and \ref{fig:distance study}; the maximum sensitivity for a device width of 15~nm is 6.9, while it is around 12.5 for a device width of 10~nm at 4~K. This enhancement in sensitivity is due to the size quantization effects in $\delta$-layers systems. A similar observation has been made in other systems, such as nanowires, in which it has been demonstrated that the sensitivity increases as the diameter of the nanowire decreases \cite{Li:2011}.

\begin{figure*}[ht!]
  \centering
  \includegraphics[width=1.0\linewidth]{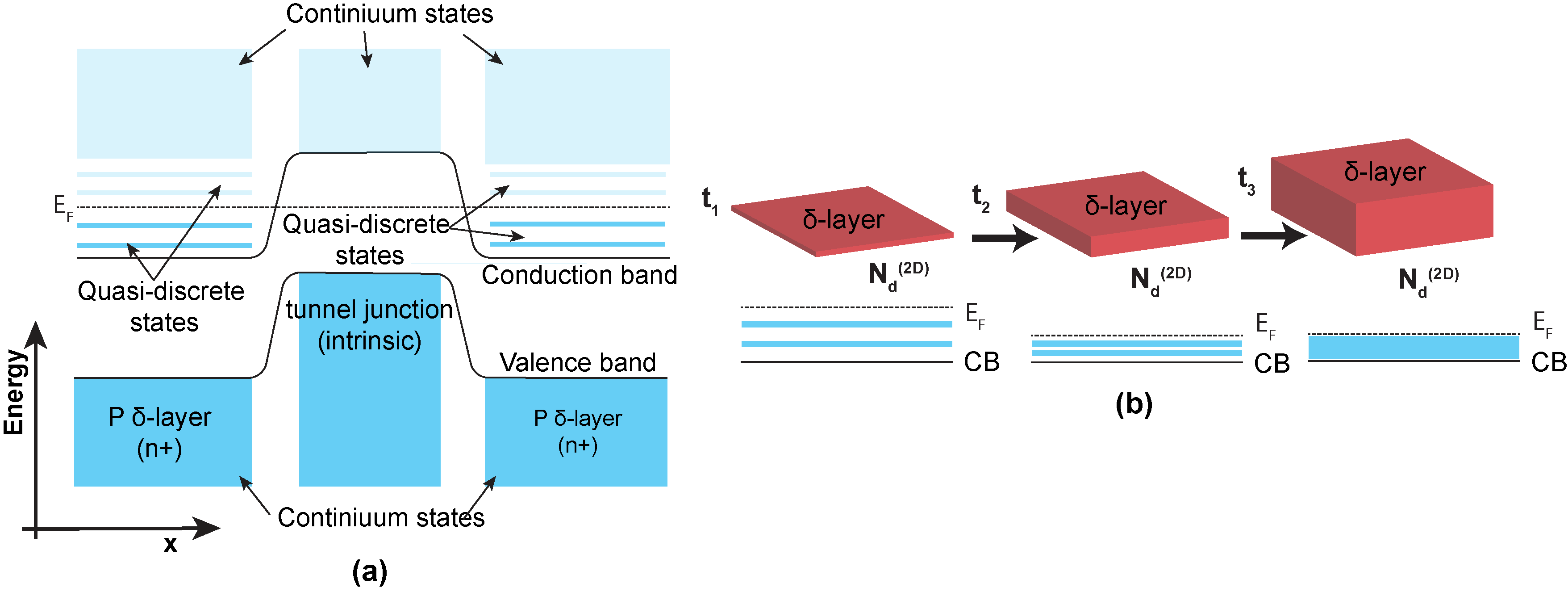}
  \caption{ (a) Schematic representation of the band structure for $\delta$-layers tunnel junctions in a semiconductor; The quantized states are represented in the figure with thin stripes in the low-energy conduction band, whereas the continuous states  are represented with thick stripes. (b) Schematic representation of the effect of the $\delta$-layer thickness on the conduction band structure.}
  \label{fig:Band structure for delta-layers}
\end{figure*}

Figure~\ref{fig:sensitivity vs thickness} shows the sensitivity of the device as a function of the thickness of the $\delta$-layer for both temperatures, 4 and 300~K, and at two applied voltages between the source and drain of 1 and 100~mV. For this study, a single (negative) charge is placed in the middle of the device, between the $\delta$-layers, at the coordinates $(x,y,z)=(L/2,H/2,W/2)$, and we evaluate how the tunnel current changes for various thicknesses. One can clearly observe from these results that the maximum sensitivity is achieved for thinner $\delta$-layers, after which the sensitivity decreases as the thickness increases.  One can also note that at higher temperatures the sensitivity is higher at lower voltages.  Thus, these results demonstrate that thinner $\delta$-layers are more sensitive than thicker ones and, as we discuss in the following, the higher sensitivity for thinner $\delta$-layers arises from the strong quantization of the conduction band in these highly-confined systems, which is the result of the confinement of the dopants in the z-direction.

In our previous work, Refs.~\citenum{Mamaluy:2021} and \citenum{Mendez_CS:2022}, we found that the conduction band structure of $\delta$-layers is strongly quantized, resulting in quasi-discrete states in the low-energy conduction band. 
A schematic representation of the conduction band quantization for $\delta$-layers is shown in Figure~\ref{fig:Band structure for delta-layers}a. The quantized states are represented in the figure with thin stripes in the low-energy conduction band, whereas the continuous states  are represented with thick stripes. The Fermi-Dirac distribution determines the occupancy of states, distinguishing between occupied and unoccupied states. 
In Refs.~\citenum{Mamaluy:2021} and \citenum{Mendez_CS:2022}, we also found that the number of quantized states, as well as the corresponding splitting energy between them, are strongly dependent on both the $\delta$-layer thickness $t$ and the doping density $N_D$. We indeed determined that as we increase the thickness of the $\delta$ layer and maintain constant the total charge (i.e., fixing constant the 2D doping density), the splitting energy between them becomes smaller, resulting in less quantization of the conduction band (Figure~\ref{fig:Band structure for delta-layers}b). These previous results, together with the results from Figure~\ref{fig:sensitivity vs thickness}, demonstrate that thinner $\delta$-layers are more sensitive than thicker ones due to the stronger quantization of the conduction band. As the thickness of the $\delta$-layer increases, the conduction band becomes less quantized. Consequently, this leads to a reduction in deviation of the tunneling rate from the ideal rate in the presence of charges, thereby decreasing the sensitivity of the $\delta$-layer to detect charges. We also note that the proposed extreme sensing mechanism using discrete energy levels is reminiscent of charge sensing using single-electron transistors (SETs)\cite{Kiyama2018}, but without the need for the Coulomb blockade effect\cite{Averin-Likharev:1986}. In $\delta$-layer systems, the quasi-discrete energy levels are located within the $\delta$-layer regions, instead of the "island", which is unnecessary for $\delta$-layer tunnel junctions, as shown in Figure~\ref{fig:Band structure for delta-layers}a.

As demonstrated in this work, APAM tunnel junctions are very sensitive to the presence of charges, making these devices very suitable for sensing applications that involve charges. Examples of these applications include, but are not limited to: (i) biomolecular and chemical sensing; (ii) radiation detection; (iii) quantum computing; (iv) rapid surface imaging with atomic resolution; (v) brain–computer interface (BCI), to unlock and interpret the brain's electrical signals using electrocorticography with high spatial resolution and potential capability to capture the smallest signals. 

\section{Conclusions}\label{sec:conclusion}

In this work, we have demonstrated the extreme sensitivity of APAM devices, based on $\delta$-layer tunnel junctions embedded in a semiconductor, for charge sensing in the low-concentration limit, i.e., for detecting single charges or low concentrations of charges. These devices have the potential to enable new opportunities in a range of applications, including biological, chemical, and radiation sensing. These devices also offer several key advantages: (i) the exceptional simplicity of a two-terminal device; (ii) extremely small dimensions, on the order of 10~nm by 10~nm, enabling the integration of a high density of elements per unit area; and (iii) the potential integration with CMOS technology. These advantages make these devices strong candidate for use in an array configuration, where signals from multiple sensing cells can be processed "on-site" using neuromorphic architecture to provide real-time analysis, enhance sensitivity and selectivity, and support the simultaneous detection of multiple target elements \cite{Maity:2023}.

Specifically, we have investigated how the tunneling current is affected by the presence of a single positive/negative charge near the tunnel junction in a two-terminal P-doped $\delta$-layer tunnel junction embedded in silicon at cryogenic and room temperature. We have found that the presence of a single charge strongly affects the tunneling current, particularly for negative charges, resulting in a high sensitivity of these devices. We have also demonstrated that these devices can enhance the sensitivity in the limit, exhibiting superior sensitivity compared to (T)FET-based sensors. The sensitivity metric, defined as the ratio of the current change to the nominal change, indicates how much the tunneling rate changes relative to the nominal current of the device. In fact, our results show that our APAM device achieves superior sensitivity: the maximum sensitivity is around $6.9$ at 4~K and $0.86$ at 300~K for a single negative charge, which is significantly higher than the theoretical upper bound of sensitivity for TFET-based sensors, typically $S=6 \times 10^{-3}$ for low concentrations of charges. As exhibited in this work, the sensitivity of these devices at low temperature can be also enhanced by reducing the width of the $\delta$-layers: for a $\delta$-layer width of 10~nm, the sensitivity increases up to approximately 12.5 at 4~K. We have also demonstrated that APAM tunnel junctions can effectively sense charges far from the tunnel junction, specifically at 4~K.  Finally, we have proposed that the extreme sensitivity to the presence of charges arises from the strong quantization of the conduction band in these highly confined systems.

\section{Methods}\label{sec:method}
The simulations in this work are conducted using the open-system charge self-consistent Non-Equilibrium Green Function (NEGF) Keldysh formalism \cite{Keldysh:1965,Datta:1997}, together with the Contact Block Reduction (CBR) method \cite{Mamaluy:2003,Mamaluy_2004,Sabathil_2004,Mamaluy:2005,Khan:2007,Khan_2008,Gao:2014,Mendez:2021} and the effective mass theory. The CBR method allows a very efficient calculation of the density matrix, transmission function, etc. of an arbitrarily shaped, multiterminal two- or three-dimensional open device within the NEGF formalism and scales linearly $O(N)$ with the system size $N$. As validation, in our previous works \cite{Mendez:2020,Mamaluy:2021}, we demonstrated a very good agreement with experimental electrical measurements for Si: P $\delta$-layer systems \cite{Goh:2006,Goh:2009,Reusch:2008,McKibbin:2014}, proving an excellent reliability of this framework to investigate $\delta$-layer systems. Similarly, our published results in Refs.~\citenum{Mendez_CS:2022} and \citenum{Mendez:2023}, without fitting parameters, agree remarkably well with the most recent experimental data for tunnel junctions in these systems\cite{Donnelly:2023}.

Within this framework, we solve charge self-consistently the open-system Schr\"{o}dinger equation and the non-linear Poisson equation \cite{Mamaluy:2003,Mamaluy:2005,Gao:2014}. Here we employ a single-band effective mass tensor approximation for the kinetic energy operator with a valley degeneracy of $d_{val}=6$. For the charge self-consistent solution of the non-linear Poisson equation, we use a combination of the predictor-corrector approach and Anderson mixing scheme \cite{Khan:2007,Gao:2014}. First, an eigenvalue problem is solved for a specially defined closed system, while taking into account the Hartree potential $\phi^H(\textbf{r}_i)$ and the exchange and correlation potential $\phi^{XC}(\textbf{r}_i)$ \cite{PerdewZunger:1981}. Second, the local density of states (LDOS) of the open system, $\rho(\textbf{r}_i,E)$, and the electron density, $n(\textbf{r}_i)$, are computed using the CBR method for each iteration. Then, the electrostatic potential and the carrier density are used to calculate the residuum $F$ of the Poisson equation
\begin{equation}
\big|\big|\textbf{F}[\boldsymbol\phi^H(\textbf{r}_i)]\big|\big|=\big|\big|\textbf{A}\boldsymbol\phi^H(\textbf{r}_i) - (\textbf{n}(\textbf{r}_i)-\textbf{N}_D(\textbf{r}_i)+\textbf{N}_A(\textbf{r}_i))\big|\big|,
\end{equation}
where $\textbf{A}$ is the matrix derived from the discretization of the Poisson equation, and $\textbf{N}_D$ and $\textbf{N}_A$ are the total donor and acceptor doping densities arrays, respectively. If the residuum is larger than a predetermined threshold $\epsilon$, the Hartree potential is updated using the predictor-corrector method, together with the Anderson mixing scheme. Using the updated Hartree potential and the corresponding carrier density, the exchange-correlation is computed again for the next step, and an iteration of the Schr\"{o}dinger-Poisson equations is repeated until the convergence is achieved with $\big|\big|\textbf{F}[\boldsymbol\phi^H(\textbf{r}_i)]\big|\big|<\epsilon=5\times 10^{-6}$~eV.
In our simulations, we have utilized a 3D real-space model, with a discretization size of 0.2~nm along all directions, thus with about $10^{6}$ real-space grid points, and up to 4,000 energy points were used. The CBR algorithm automatically ascertains that out of more than 1,000,000 eigenstates only about 1000 ($<0.1\%$) of lowest-energy states is needed, which is generally determined by the material properties (e.g., doping level) and the temperature of the system, but not the device size. We have also employed the standard values of the inertial effective mass tensor for electrons, $m_l = 0.98 \times m_e$, $m_t = 0.19 \times m_e$, and the dielectric constant of silicon $\epsilon_{Si}=11.7$.  Further details of the methodology to study these systems can be found in our previous publications \cite{Mamaluy:2021,Mendez:2021,Mendez_CS:2022,Mendez:2023b}.

The computational device used in this work, shown in Figure~\ref{fig:APAM and computational_model}c, consists of a semi-infinite source and drain, in contact with the device of length $L$. By using the NEGF open boundary conditions, the source and drain represent a way to extend the device into infinity along the x-axis. The device is composed of a lightly doped Si body and Si cap and two very thin, highly n-type (e.g., P) doped layers separated by an intrinsic gap of length $L_{gap}$. 
We selected the silicon crystallographic direction [010] along the propagation direction (i.e., the x-axis), and [100] as the direction perpendicular to the $\delta$-layer (i.e., along the z-axis). While the crystal orientation may affect the sensitivity magnitude, this aspect lies beyond the scope of the present study.
If not specified otherwise in the manuscript, the device length $L$ is set to $30~\text{nm} + L_{gap}$ to avoid  boundary effects between the source and drain contacts, the tunnel gap length $L_{gap}$ is 10~nm, the device height $H$ is $8$~nm, and the device width $W$ is $15$~nm, with an effective width of $13$~nm for the $\delta$-layers, to minimize  size quantization effects on the conductive properties\cite{Mendez_CS:2022}. Similarly, we have considered a thickness of $\delta$-layer of $t=1$~nm, a sheet doping density of $N_D=10^{14}$~cm$^{-2}$ ($N_D^{(2D)} = t \times N_D^{(3D)}$) in the $\delta$-layers, and a doping density of $N_A=5\times 10^{17}$~cm$^{-3}$ in the Si cap and Si body. These doping densities and dimensions are of the order of published experimental work \cite{Ward:2020,Donnelly:2023,Skeren:2020,McKibbin:2009,Polley:2012}. In our simulations, the negative/positive charge is modeled  by approximating a point charge with a density of (positive or negative) $4.6 \times 10^{21}$~cm$^{-3}$ homogeneously distributed in a total volume of (0.6~nm)$^3$. The final spatial distribution of the total charge will be dictated by the self-consistent solution of the open-system Schr\"{o}dinger and Poisson equations.  Finally, we have neglected inelastic scatterings in our simulations. At the cryogenic temperature of 4.2~K, inelastic scattering can be neglected \cite{Goh:2006,Mazzola:2014} since phonon are effectively freeze-out; at room temperature, we assume than the impact of thermionic emission on the sensitivity is greater than that of inelastic scattering. However, further studies need to be done to elucidate the role of inelastic scattering on electron transport at room temperature.

\section*{Data availability}
The datasets used and/or analyzed during the current study are available from the corresponding authors on reasonable request.

\section*{Acknowledgment}\label{sec:acknowledgement}
All authors gratefully acknowledge funding support from the Department of Energy’s Advanced Simulation and Computing (ASC) Program. The authors are deeply grateful to John Wagner for his encouragement and support of this work. Sandia National Laboratories is a multimission laboratory managed and operated by National Technology and Engineering Solutions of Sandia, LLC., a wholly owned subsidiary of Honeywell International, Inc., for the U.S. Department of Energy’s National Nuclear Security Administration under contract DE-NA-0003525. This paper describes objective technical results and analysis. Any subjective views or opinions that might be expressed in the paper do not necessarily represent the views of the U.S. Department of Energy or the United States Government.

\bibliography{main}

%\section*{Author contributions}
%J.P.M. and D.M. performed equally the central calculations and analysis presented in this work.

\section*{Competing interests}
The authors declare no competing interests.

\end{document}